\documentclass[aps,pra,reprint,superscriptaddress,nofootinbib]{revtex4-1}

\usepackage{amsmath}
\usepackage{bbold}
\usepackage{amsfonts}
\usepackage{graphicx}
\usepackage{verbatim}
\usepackage{tikz}

\newcommand{\ket}[1]{\, | #1 \rangle}

\begin{document}
\title{Creation and transfer of non-classical states of motion using Rydberg dressing of atoms in a lattice}

\author{L.~F.~Buchmann}
\affiliation{Department of Physics and Astronomy, Aarhus University, Ny Munkegade 120, DK 8000 Aarhus C, Denmark}
\affiliation{Institute of Electronic Structure and Laser, FORTH, GR-71110 
Heraklion, Crete, Greece}
\author{K.~M\o lmer}
\affiliation{Department of Physics and Astronomy, Aarhus University, Ny Munkegade 120, DK 8000 Aarhus C, Denmark}
\author{D.~Petrosyan}
\affiliation{Institute of Electronic Structure and Laser, FORTH, GR-71110 
Heraklion, Crete, Greece}

\begin{abstract}
We theoretically investigate the manipulation of the motional states 
of trapped ground-state atoms using Rydberg dressing via nonresonant 
laser fields. The forces resulting from Rydberg-state interaction between 
dressed neighboring atoms in an array of microtraps or an optical lattice 
can strongly couple their motion. We show that intensity modulation of 
the dressing field allows to squeeze the relative motion of a pair of atoms 
and generate non-classical mechanical states. Extending this pairwise scheme 
to one-dimensional chains provides flexible control over the mechanical 
degrees of freedom of the whole system.  We illustrate our findings with 
protocols to manipulate all motional degrees of freedom of a pair of atoms 
and create entangled states. We also present a method to transfer 
non-classical correlations along an atomic chain of nontrivial length. 
The long-lived nature of motional states, together with the high tunability 
of Rydberg dressing, makes our proposal feasible for current experimental 
setups.
\end{abstract}

\maketitle

\section{Introduction}
\label{sec:intro}

In atomic physics, spatial degrees of freedom are often seen solely as sources of decoherence and dephasing~\cite{lesanovsky2013, Turchette2000}, which can be reduced to their fundamental limits thanks to ever-improving techniques to cool and control atoms. This simplifies the treatment of the internal degrees of freedom of atoms which are used to perform various tasks, e.g., in quantum computation~\cite{deutsch1999, david2014} and simulation~\cite{qsim} or quantum-enhanced sensing~\cite{haroche2016}. Additionally, simplified treatments are frequently justified by the lack of sensitivity of many experiments to positions and momenta of atoms. Recent experiments have demonstrated increased spatial resolution of individual atoms via coupling their motion to the field of an optical resonator~\cite{Esslinger2010, stamperkurn2010,stamperkurn2013} or through the realization of quantum gas microscopes~\cite{greiner2009, bloch2010}. This increased experimental sensitivity lays the groundwork for the exploitation of atomic motion as a quantum resource, as is successfully done in trapped ion quantum computing~\cite{ionqc}.

The large separation of frequencies and coupling strengths of atoms to the phononic and photonic environments makes their mechanical degrees of freedom long lived compared to the radiative ones. This renders atomic motion an attractive candidate for quantum information storage, quantum computation or simulation \cite{Lloyd1999}. Manipulation of a large number of quantum harmonic oscillators to engineer the state of the entire system would provide a powerful tool for quantum technologies, including sensing, computation and simulations~\cite{singh2016, poletti2016}. Ultracold atomic samples trapped in an optical lattice or array of microtraps provide a large number of quantum harmonic oscillators and Rydberg-state dressing~\cite{pohl2014, pfau2014} allows controlled couplings between individual oscillators. 

Driving a ground-state atom with a laser far detuned from a Rydberg state leads to a small fraction of the atomic wave function to occupy that highly excited state. The dressed atoms inherit some of the Rydberg-state properties, namely, strong, long-range interactions between the Rydberg excited atoms \cite{RydAtoms,rydQIrev,rydDBrev}. At the same time, decoherence due to the radiative decay is suppressed as a result of the small occupation probability of the excited state. A pair of dressed atoms may then experience strong, spatially dependent interactions which can be tuned via the dressing laser. A proper choice of the Rydberg state and the dressing laser parameters can then provide flexible control of the mechanical degrees of freedom of the system. Rydberg dressing has been demonstrated experimentally via spectroscopic means in individually trapped atoms~\cite{deutsch2015} and in optical lattices~\cite{bloch2016}. 

In this article, we exploit the atomic motion as a coherent quantum degree of freedom. Rydberg dressing allows to manipulate the mechanical state of a system of many trapped neutral atoms. Non-classical correlations can be established between neighboring atoms and distributed controllably within the system. Compared to schemes based on optical resonators to mediate the coupling of atomic motion, Rydberg dressing is less susceptible to intrinsic decoherence~\cite{buchmann2015, spethmann2016}. Using experimentally feasible parameters, we demonstrate the possibilities of our proposal with three conceptually simple but important applications: the extension of the accessible degrees of freedom via auxiliary atoms, the creation of highly entangled states and the transfer of non-classical correlations along a chain of neutral atoms.

\section{Effective Interaction}
\label{sec:effint}

\begin{figure}[t]
\begin{center}
\includegraphics[width=0.8\columnwidth]{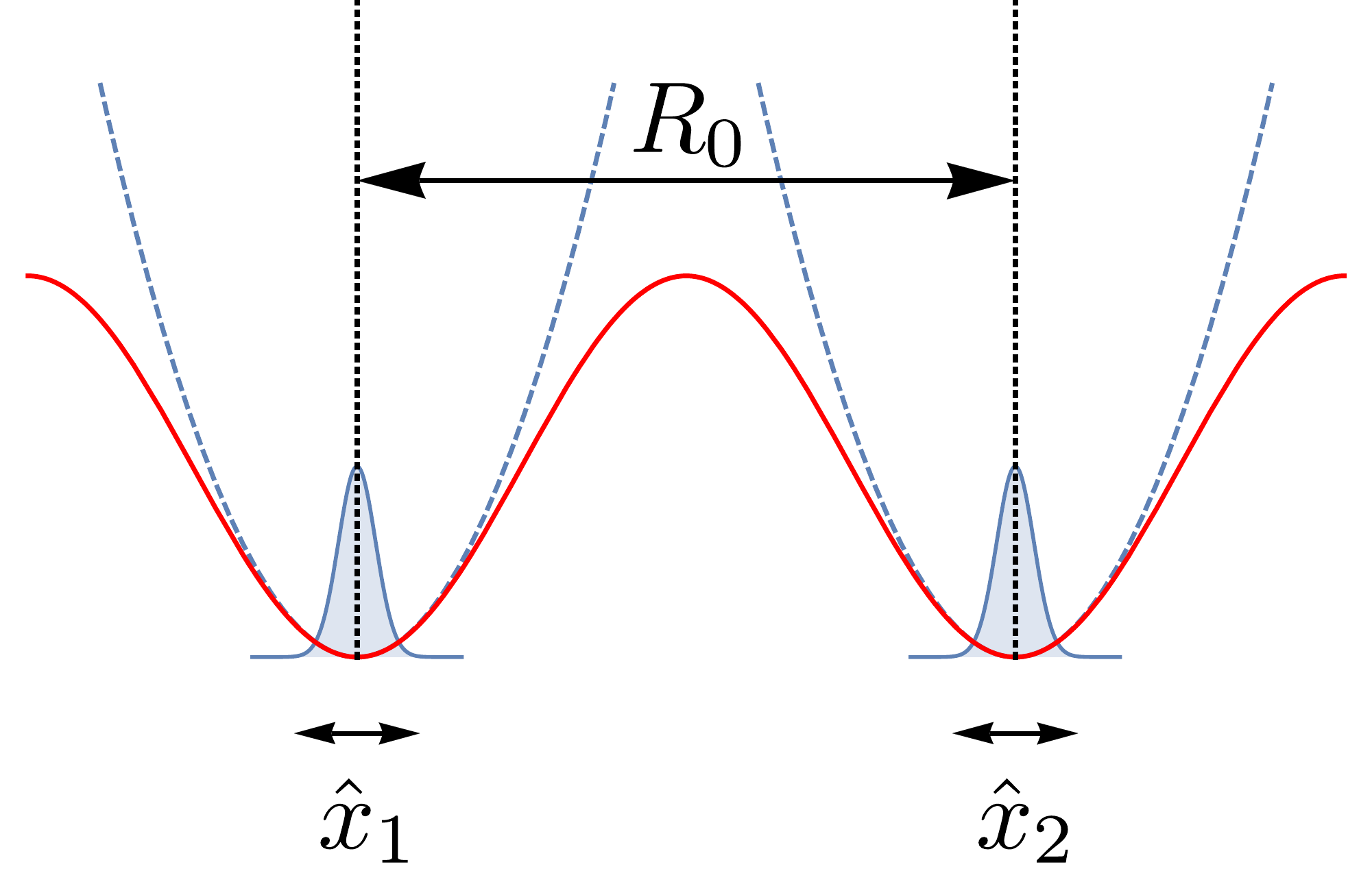}
\end{center}
\caption{Sketch of the considered setup. Two atoms are confined in individual, approximately harmonic, traps whose minima are separated by distance $R_0$.}
\label{sketch}
\end{figure}

\begin{figure}
\includegraphics[width=0.5\textwidth]{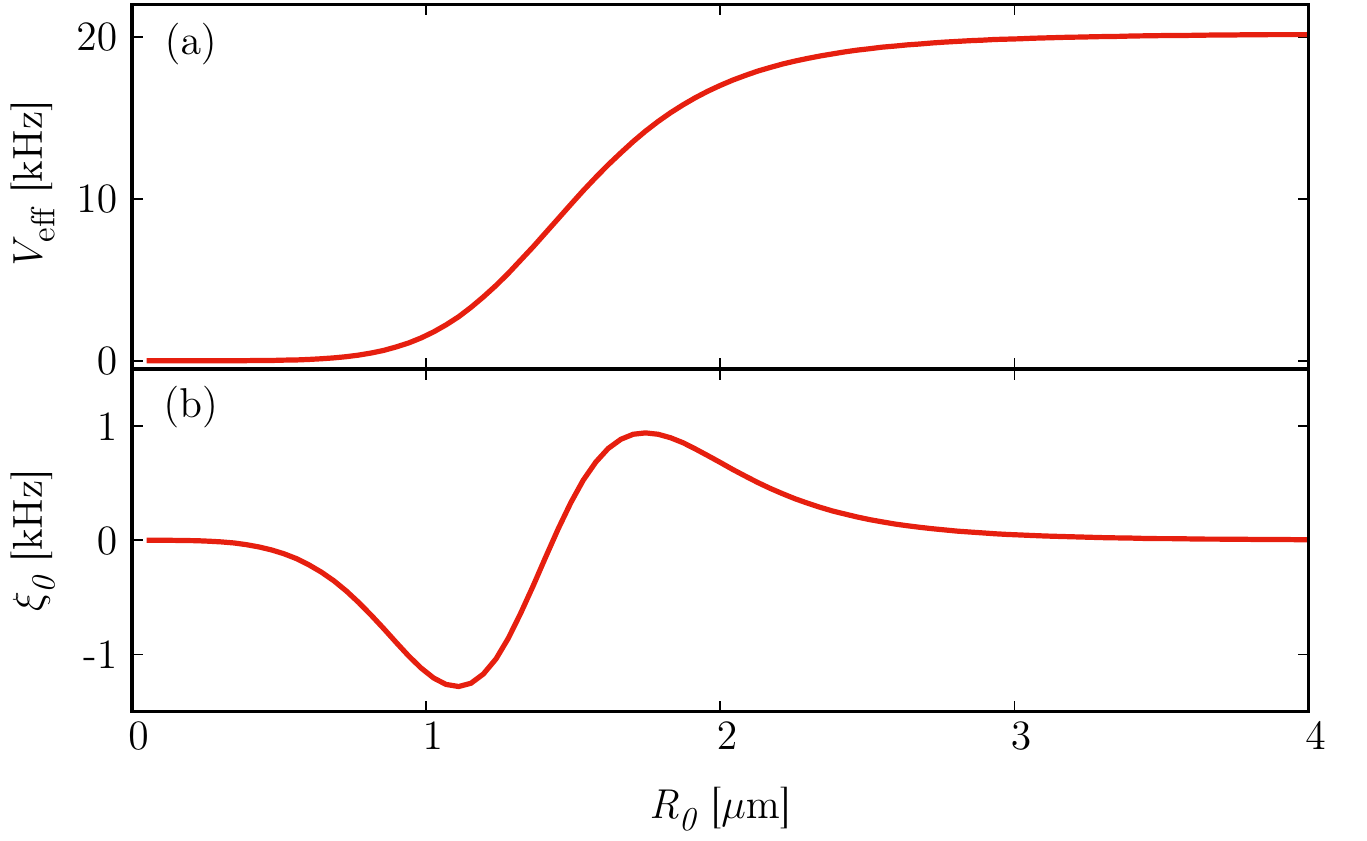}
\caption{The effective interaction potential (top) and coupling strengths (bottom) for a pair of dressed atoms, in units of kHz. 
We assume ${}^{87}$Rb atoms in a trap with oscillation frequency $\omega_m/2\pi=2$kHz, and the $n=53$ Rydberg state~\cite{browaeys2013} dressed with a laser of Rabi frequency $\Omega/2\pi=10$MHz and detuning $\Delta=10 \Omega$.}
\label{interactionstrengths}
\end{figure}

We consider two neutral atoms in individual harmonic traps with minima separated by distance $R_0$, as sketched in Fig.~\ref{sketch}. 
A strong laser field is driving the transition from the trapped
ground state $\ket{g}$ to an untrapped Rydberg state $\ket{e}$ with Rabi frequency $\Omega(t)$ and detuning $\Delta$. The Hamiltonian governing the evolution of the electronic states is
\begin{align}
H=&
-\Delta\Big[\mathbb{1}\otimes|e\rangle\langle e|+|e\rangle\langle e|\otimes\mathbb{1}\Big]
\nonumber\\
&+\Omega(t)\Big[\mathbb{1}\otimes\big(|g\rangle\langle e|+|e\rangle\langle g|\big)
+\big(|g\rangle\langle e|+|e\rangle\langle g|\big)\otimes\mathbb{1}\Big]
\nonumber\\
&
+V(R)|e\rangle\langle e|\otimes|e\rangle\langle e|,
\label{hamiltonian}
\end{align}
with $V(R)$ the interaction potential between the pair of atoms in the Rydberg state. We assume that the driving laser is far detuned from any transition in the system, such that states containing Rydberg excitations can be adiabatically eliminated. After the elimination, see appendix~\ref{appa} or \cite{bloch2016, pfau2014} for details, we arrive at the effective interaction potential for two ground-state atoms
\begin{align}\label{helim}
V_\mathrm{eff}=\frac{2|\Omega(t)|^2}{\Delta}+\frac{4|\Omega(t)|^4/\Delta}{2(\Delta^2-|\Omega(t)|^2)-\Delta V(R)}.
\end{align}
Here the first term corresponds to the dispersive (ac Stark) shifts of the ground states of individual atoms due to virtual excitation. This term is a $V(R)$ independent c-number acting trivially on spatial wavefunctions and we can neglect it. The second term describes the consequences of the interaction of the pair of atoms via the Rydberg states. It leads to a spatially dependent evolution of the phase of the system wavefunction and influences the mechanical behavior of the two atoms. For Rydberg states that interact via the van der Waals potential, we have 
\begin{align}\label{force}
V(R)= \frac{C_6}{|R_0+\hat{x}_1-\hat{x}_2|^6}\equiv V(\hat{x}_1,\hat{x}_2),
\end{align}
where $R_0$ is the equilibrium separation of the two atoms and we have introduced operators $\hat{x}_i$ describing the displacement of each atom from its equilibrium position along the axis connecting the two atoms. For ultracold atoms located in separate optical traps, we assume \mbox{$R_0\gg\sqrt{\langle\hat{x}_i^2\rangle}$}. The dressed interaction potential is obtained by substituting $V(\hat{x}_1,\hat{x}_2)$ into Eq.~(\ref{helim}):
\begin{align}
V_\mathrm{eff}(\hat{x}_1,\hat{x}_2)=\frac{4|\Omega(t)|^4/\Delta}{2(\Delta^2-|\Omega(t)|^2)-\Delta V(\hat{x}_1, \hat{x}_2)}.
\end{align}
This effective potential exhibits the experimentally confirmed soft-core character~\cite{bloch2016} typical of Rydberg dressing, see Fig.~\ref{interactionstrengths}(a). Detailed calculations of the interaction potential including more states result in quantitative differences and more complicated behavior at short distances. Such calculations, however, do not lead to qualitatively different results at larger interatomic separation that we are interested in~\cite{bloch2016, pfau2014}. The two main ingredients that allow coherent manipulation of the atomic motional states are the non-linearity of the interaction and the absence of further resonances. Both of these features are present in our simplified description which is justified in the regimes we consider. 

We can now write the effective Hamiltonian describing the mechanical behavior of the system, assuming harmonic trapping of both atoms with equal trapping frequencies $\omega_m$: 
\begin{align}
H_\mathrm{eff}=\sum_{i=1}^2H_m^{(i)}+V_\mathrm{eff}(\hat{x}_1,\hat{x}_2),
\end{align}
with
\begin{align}\label{hmot}
H_m^{(i)}=\frac{\hat{p}_i^2}{2m}+\frac{m\omega_m^2}{2}\hat{x}^2_i,
\end{align}
where $m$ is the atomic mass and $\hat{p}_i$ are momentum operators conjugate to $\hat{x}_i$. 

Expanding the effective potential to second order in the relative displacement $\delta\hat{x}=\hat{x}_1-\hat{x}_2$ leads to
\begin{align}\label{effhamiltonian}
H_\mathrm{eff}=\sum_{i=1}^2H_m^{(i)}-J(t)|\Omega(t)|^4\delta\hat{x}-\xi(t)|\Omega(t)|^4\delta\hat{x}^2, 
\end{align}
where we have dropped a constant and introduced 
\begin{align}
J(t)
=&\frac{24C_6/R_0^7}{\left(\Delta(V(R_0)-2\Delta)+2|\Omega(t)|^2\right)^2},
\\
\xi(t)
=&\frac{12C_6/R_0^{8}(\Delta(5V(R_0)+14\Delta)-14|\Omega(t)|^2)}{(\Delta(V(R_0)-2\Delta)+2|\Omega(t)|^2)^3}.
\end{align}
In the far off-resonant regime $\Delta\gg|\Omega(t)|$ we may take the leading non-vanishing order in $\Omega(t)/\Delta$ in these expressions, obtaining
\begin{align}
J=&\frac{24C_6/R_0^7}{\Delta^2(V(R_0)-2\Delta)^2},
\\
\xi
=&\frac{12C_6/R_0^{8}(5V(R_0)+14\Delta)}{\Delta^2(V(R_0)-2\Delta)^3}.
\end{align}
Note that we neglect terms of the order $\langle\hat{x}_1-\hat{x}_2\rangle^3/R_0^3$ and higher~\footnote{If the equilibrium distance $R_0$ is not large enough to justify dropping these terms by their magnitude, there is a non-zero detuning which makes the coefficient of the next order term vanish exactly.}.

The term proportional to $J$ in Eq.~(\ref{effhamiltonian}) cannot couple the atomic motion in a non-factorizable manner because it is separable with respect to the two atoms. This term can be used as a drive to pump the system into a coherent state of motion. Because the Hamiltonian $H_\mathrm{eff}$ is quadratic in the position and momentum operators, the first and second moments of $\hat{x}_i$ and $\hat{p}_i$ decouple and the expectation values of single operators follow classical equations of motion. We do not consider these dynamics here but instead focus on the second moments describing the fluctuations of the atomic motional degrees of freedom. Consequently, for the remainder of this paper, we can neglect the term proportional to $J$.

On the other hand, in Eq.~(\ref{effhamiltonian}), the term proportional to $\xi$ acts as an effective spring between the two atoms, shifting their trapping frequencies and coupling their motion. Modulating the intensity of the dressing laser allows to tune the spring constant in time and realize a parametric oscillator. 

To clarify our treatment, we write $\hat{x}_j=x_{\mathrm{zpt}}(\hat{a}_j+\hat{a}_j^\dag)$ and $\hat{p}_j=i p_{\mathrm{zpt}}(\hat{a}^\dag_j-\hat{a}_j)$, where $x_{\mathrm{zpt}}$ and $p_{\mathrm{zpt}}$ are the zero point widths in position and momentum for the trapped atoms in the absence of the dressing laser. Absorbing $x_{\mathrm{zpt}}$  into the coupling via
$\xi\to x_{\mathrm{zpt}}^2 \xi$ results in the Hamiltonian
\begin{align}
H=&\sum_i\left[(\omega_m-2\xi|\Omega(t)|^4)\hat{a}_i^\dag\hat{a}_i-\xi|\Omega(t)|^4(\hat{a}_i^2+(\hat{a}_i^\dag)^2)\right]
\nonumber\\
&+2\xi|\Omega(t)|^4(\hat{a}_1\hat{a}_2+\hat{a}_1^\dag\hat{a}_2^\dag+\hat{a}_1^\dag\hat{a}_2+\hat{a}_1\hat{a}_2^\dag).
\end{align}

Consider now dressing fields with temporal behavior
\begin{align}\label{omegatimedep}
\Omega(t)=\bar{\Omega}(t)(1+A(t)\cos(\omega_dt)),
\end{align}
where $0\le A(t) \le 1$ is the modulation depth, with the time dependence of both $\bar{\Omega}(t)$ and $A(t)$ given by step functions. Such driving leads to the spring coefficient
$\propto |\Omega(t)|^4$ oscillating at four different frequencies. In a regime where $\bar{\Omega}\ll\omega_m$, we may neglect off-resonant terms within the rotating wave approximation, see Appendix \ref{appb} for details. In the frame rotating at the unperturbed trapping frequency $\omega_m$, we then have 
\begin{align}
\label{inthamiltonian}
H=H_0+H_1,
\end{align}
where
\begin{subequations}\label{explicithamiltonian}
\begin{align}
H_0=&-\xi_0\sum_j\hat{a}_j^\dag\hat{a}_j+\xi_0\left(\hat{a}_1^\dag\hat{a}_2+\hat{a}_1\hat{a}_2^\dag\right),
\\
H_1=&-\frac{\xi_1}{2}\sum_j\left(e^{-2i\xi_0 t}\hat{a}_j^2+e^{+2i\xi_0 t}(\hat{a}_j^\dag)^2\right)
\nonumber\\
&+\xi_1\left(e^{-2i\xi_0 t}\hat{a}_1\hat{a}_2+e^{+2i\xi_0 t}\hat{a}_1^\dag\hat{a}_2^\dag\right),
\end{align}
\end{subequations}
with $\xi_0=\xi|\bar{\Omega}|^4f_0$ and $\xi_1=\xi|\bar{\Omega}|^4f_1$, where $f_{0,1}$ are dimensionless parameters of order unity (see Appendix \ref{appb}).
Spring shifts and a beam-splitter like interaction between the oscillators are contained in $H_0$, whose terms are always resonant. Parametric terms, which only influence the dynamics for $A(t)\neq 0$, are in $H_1$. We plot $\xi_0$ as a function of the equilibrium interatomic distance $R_0$ in Fig.~\ref{interactionstrengths}(b). The magnitude and sign of the interaction may be tuned through $R_0$. As discussed below, the strength of the interaction may be chosen large enough to overcome motional decoherence, while the rotating wave approximation remains valid. 

We describe the motion of both atoms by the operator vector $\hat{\bf{a}}=\left(\hat{a}_1,\hat{a}_1^\dag,\hat{a}_2,\hat{a}_2^\dag\right)^\top$, where $\cdot^\top$ denotes transposition. Dressing the atoms with an intensity-modulated laser for a time $t_1$ changes the state of the system according to
\begin{align}
\hat{\bf{a}}(t_1)=\mathrm{P}(t_1)\hat{\bf{a}}(t_0),
\label{poststate}
\end{align} 
where the matrix $\mathrm{P}(t_1)$ can be found by exponentiating the coupling matrix of the Heisenberg equations of motion (see Appendix \ref{appc}).
The depth of the modulation determines the dominant character of the dynamics. If no modulation is present, $A(t)=0$, the interaction yields the beam-splitter like transformation 
\begin{subequations}\label{reduceddyn}
\begin{align}
\hat{a}_1(t)=&e^{i\xi_0t}\left[\cos(\xi_0t)\hat{a}_1(0)-i\sin(\xi_0t)\hat{a}_2(0))\right],\\
\hat{a}_2(t)=&e^{i\xi_0t}\left[-i\sin(\xi_0t)\hat{a}_1(0)+\cos(\xi_0t)\hat{a}_2(0)\right].
\end{align}
\end{subequations}
If the modulation is present, the interaction takes on a parametric character and the relative motion between the two atoms becomes entangled. For $A(t)>0.125$, the interaction turns into parametric amplification of the relative motion. The ratio $\xi_1/\xi_0$ may be chosen freely from 0 to 1.6 (see Appendix \ref{appc}). To keep our notation brief, we choose a modulation depth $A(t)=0.29$, resulting in $\xi_1=\xi_0\equiv\xi_P$ and the effect of the modulated dressing on the mechanical state of the two atoms is given by
\begin{subequations}\label{paramtransf}
\begin{align}
\hat{a}_1(t)=&\frac{1}{2}(1+\alpha(t)+\beta(t))\hat{a}_1(0)-\beta(t)\hat{a}_1^\dag(0)
\nonumber\\
&+\frac{1}{2}(1-\alpha(t)-\beta(t))\hat{a}_2(0)+\beta(t)\hat{a}_2^\dag(0),
\\
\hat{a}_2(t)=&\frac{1}{2}(1-\alpha(t)-\beta(t))\hat{a}_1(0)+\beta(t)\hat{a}_1^\dag(0)
\nonumber\\
&+\frac{1}{2}(1+\alpha(t)+\beta(t))\hat{a}_2(0)-\beta(t)\hat{a}_2^\dag(0),
\end{align}
\end{subequations}
with the coefficients
\begin{align}\label{alphabeta}
\alpha(t)=e^{i\xi_Pt}\cosh(\sqrt{3}\xi_Pt),\\
\beta(t)=\frac{ie^{i\xi_Pt}}{\sqrt{3}}\sinh(\sqrt{3}\xi_Pt).
\end{align}
The Hamiltonian $H$ conserves the Gaussian character of motional states and the quantum state of the system can be completely characterized by the co-variance matrix
\begin{align}
E_{ij}(t)=\frac{1}{2}\langle\{ \hat{\bf a}(t)_i,\hat{\bf a}(t)_j\}\rangle.
\end{align} 
The dynamics of $E$ are found from Eq.~(\ref{poststate}) 
\begin{align}\label{covmatrix}
E(t_1)=&
P(t_1)\cdot E(t_0)\cdot P(t_1)^\top.
\end{align}
The matrix $P(t_1)$ contains all the information about changes in the motional state of the two atoms due to the interaction; in particular, it contains all the information about the non-classicality of their correlations.

To establish non-classical correlations between position fluctuations of two atoms, we shine a dressing laser with modulated intensity on the corresponding pair. For a system initialized in its motional ground state, the only non-vanishing correlations are
\begin{align}\label{vacuumcorr}
\langle\hat{a}_j(0)\hat{a}_j^\dag(0)\rangle=1.
\end{align}
The state of the system after turning on the modulated dressing for a time $t_1$ is given by Eqs. (\ref{paramtransf}) with $t=t_1$. Such an interaction squeezes one quadrature of the relative motion below its vacuum value. The non-classical nature of the correlation is captured by the violation of the Cauchy-Schwarz inequality
\begin{align}\label{cauchyschwarz}
C_{ij}=|\langle\hat{a}_i\hat{a}_j\rangle|^2-\langle\hat{a}_i^\dag\hat{a}_i\rangle\langle\hat{a}_j^\dag\hat{a}_j\rangle\leq 0.
\end{align}
We have
\begin{subequations}\label{correlations}
\begin{align}\label{autoviol}
\langle\hat{a}_1^\dag\hat{a}_1(t_1)\rangle
=\langle\hat{a}_2^\dag\hat{a}_2(t_1)\rangle
&=2|\beta(t)|^2
\end{align}
and
\begin{align}\label{a1a2viol}
\langle \hat{a}_1\hat{a}_2(t_1)\rangle
&=(\alpha+\beta)\beta.
\end{align}
\end{subequations}
The Cauchy-Schwarz inequality now reads
\begin{align}
|\alpha+\beta|^2|\beta|^2-4|\beta|^4\le 0.
\end{align}
For $t_1>0$, we have $\beta\neq 0$ and can thus divide by $|\beta|^2$ to find
\begin{align}
|\alpha+\beta|^2-4|\beta|^2\le 0.
\end{align}
Expanding the expression on the left hand side and inserting $\alpha$ and $\beta$ from (\ref{alphabeta}) gives
\begin{align}
1+2\text{Re}\left(\alpha\beta^*\right)\le 0,
\end{align}
which is violated because $\alpha\beta^*$ is purely imaginary. 

For a pair of atoms, Rydberg-state dressing permits complete control of the first and second moments of their relative motion within the state space of pure Gaussian states. Crucially, this control includes squeezing, which provides an irreducible quantum resource~\cite{braunstein2005}. Let us now explore the possibilities that arise from extending the number of atoms.

\section{Three and more atoms}
In a typical experiment with atoms in an optical lattice~\cite{pohl2014, pfau2014, bloch2016, saffman2013} or an array of microtraps~\cite{browaeys2004}, the number of available trapping sites is much larger than two. The possibility to turn on  at any time the desired interaction between any two neighboring sites with broadly tunable strength allows us to exploit many atoms and to accomplish intricate engineering of the motional state of the system. We illustrate this versatility with three examples which highlight different aspects of the high degree of control. For these examples we assume that the system is initialized in its motional ground state, such that its correlations are given by Eq. (\ref{vacuumcorr}). 

\subsection{Auxiliary atom}
The tunable interaction between a pair of atoms may correlate their relative motion and change their individual state, but it leaves their collective center of mass motion undisturbed. This is because the underlying force, described by Eq. (\ref{force}), depends only on the separation between the two atoms. 
A physical pair of harmonic oscillators therefore reduces to a single controllable oscillator of their relative motion. This limitation may be remedied by supplementing the pair of atoms with a third atom and treating one of them as an auxiliary. A complete control theory of this set-up is beyond the scope of the present article and will be given elsewhere~\cite{futurework}. 

\begin{figure}
\includegraphics[width=0.8\columnwidth]{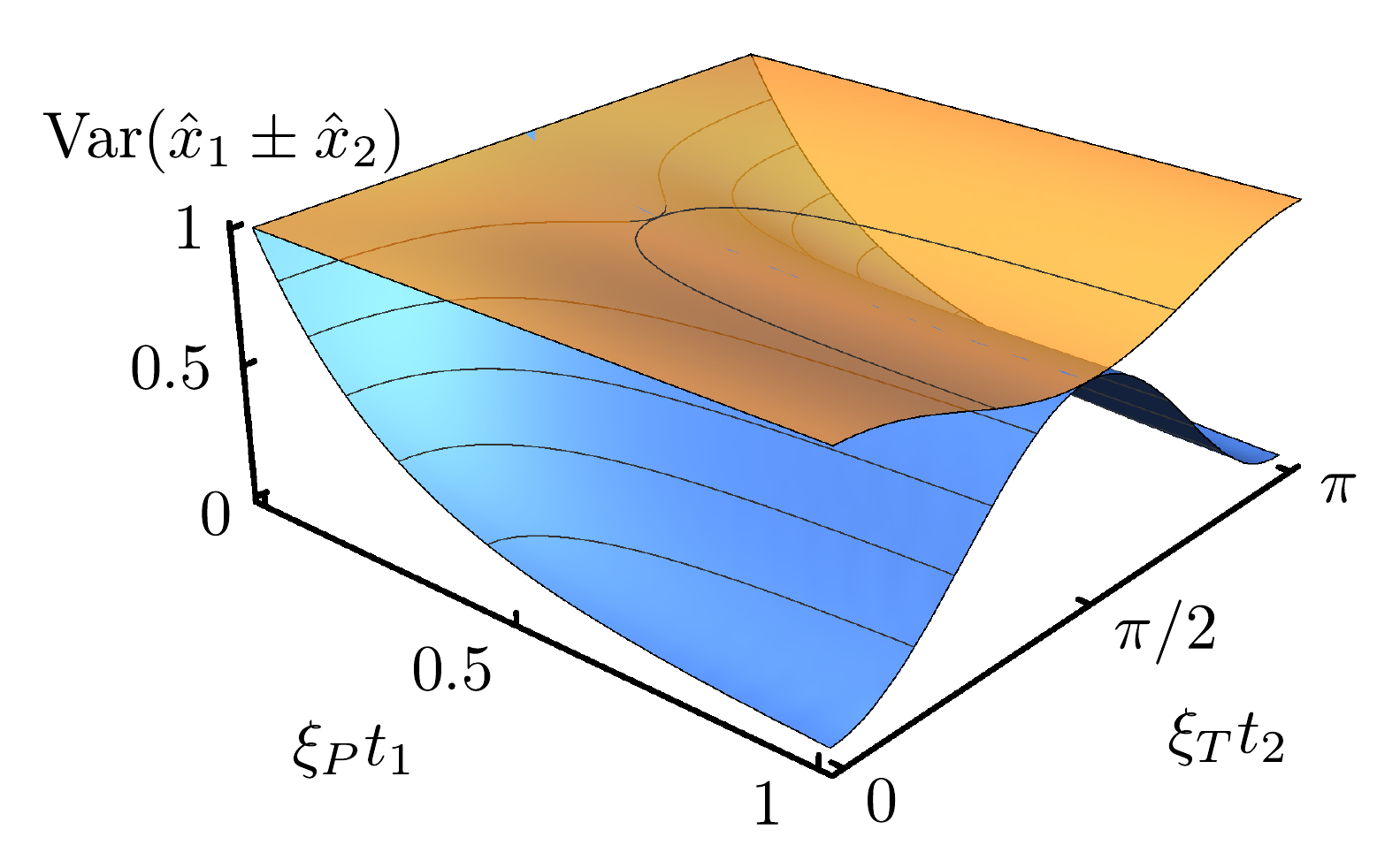}
\caption{Variance of the two collective modes of motion of two atoms 1 and 2 after squeezing their relative motion with a pulse of area $\xi_Pt_1$ and coupling one of them to an auxiliary atom with a beam-splitter like pulse of area $\xi_Tt_2$. The top and lower surfaces give the variance of the center of mass and relative mode respectively.}
\label{surfaceplot}
\end{figure}

Consider a chain of three atoms, with the last one being the auxiliary. Let us couple atoms 1 and 2 with a modulated dressing field. This field squeezes the motion of each atom as well as the relative motion of the two atoms, leaving them both in non-classical states. The squeezing can now be transferred to the center of mass mode of atoms 1 and 2 by coupling atom 2 to the auxiliary atom via the beam-splitter interaction with a strength $\xi_T$ and duration $t_2$. The variance of the center of mass mode as a function of entangling pulse area $\xi_Pt_1$ and the coupling pulse area $\xi_Tt_2$ is plotted in Fig.~\ref{surfaceplot}. 

The squeezing of the center of mass mode reaches its maximum for $\xi_Tt_2=\pi/2$. At this stage, the second and third atoms have exchanged their state and the diminished variances are simply a consequence of the non-classical nature of the motional state of atom 1.  Despite squeezed variances, the first two atoms are not entangled and quantum correlations are found only between the first atom and the auxiliary. We can improve on that situation by applying another modulated pulse with a different pulse area and squeezing phase instead of the simple beam-splitter. 

\begin{figure}[t]
\includegraphics[width=0.8\columnwidth]{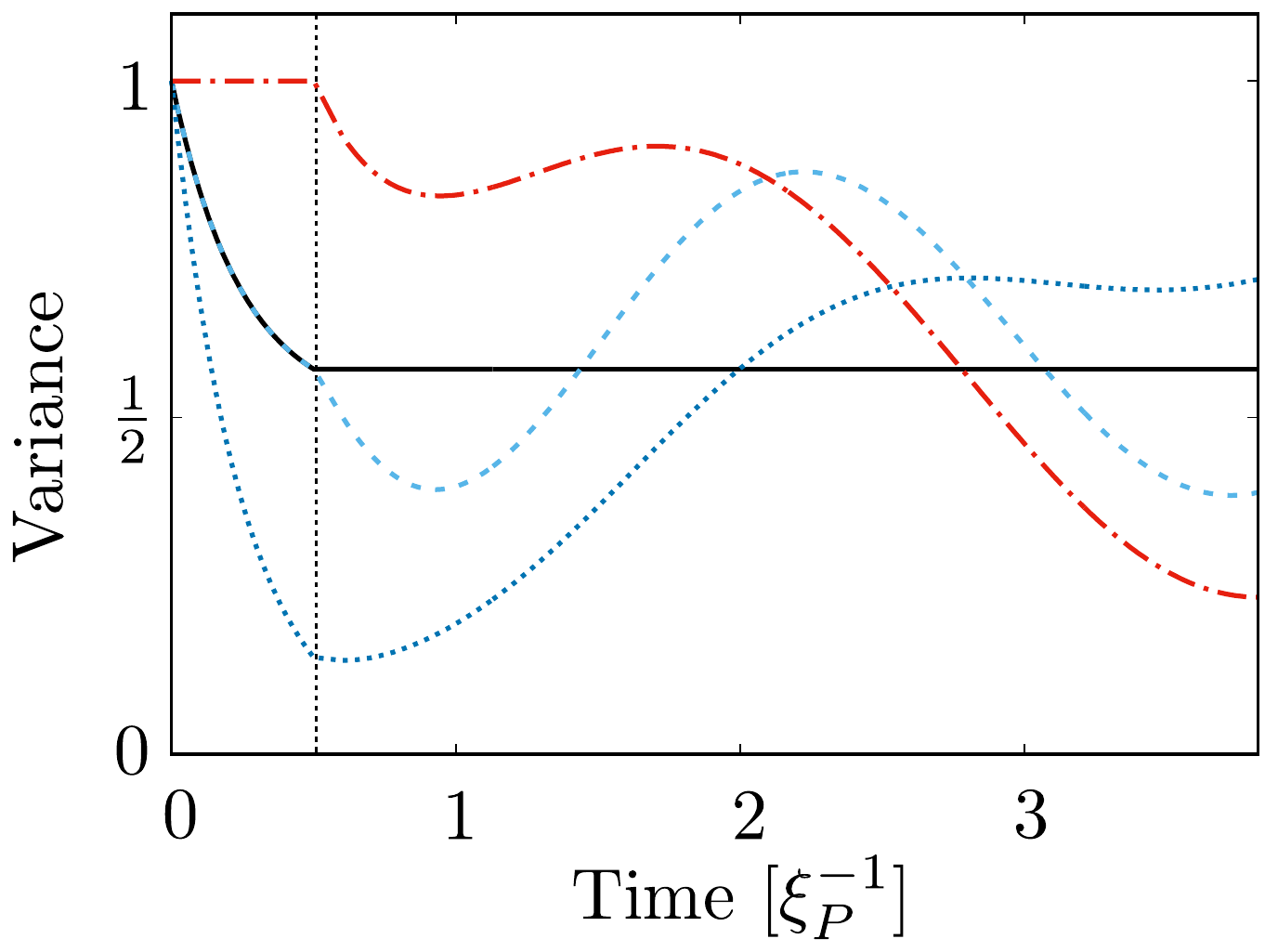}
\caption{Variances of the most squeezed quadratures of the individual atoms, 
1 (solid black) and 2 (dashed light-blue), the relative motion (dotted blue) 
and the center of mass motion (dot-dashed red) of atoms 1 and 2, 
for a two-pulse sequence optimized to squeeze the variance of the 
center of mass motion.  
Both pulses have equal strength $\xi_P$. The initial pulse squeezing the 
relative motion lasts until the dotted line, $\xi_Pt=0.5$. The second pulse 
parametrically amplifies the relative motion of atom 2 and the auxiliary. 
All variances are normalized to their vacuum value.}
\label{optvariances}
\end{figure}

We have analytic equations for the final state of the system, Eq. (\ref{covmatrix}), and may treat the area and phase of the second pulse as free parameters to achieve any chosen correlation. The resulting equations are, however, transcendental (see appendix \ref{appc}), which prevents us from making general statements about the exact controllability of the system. We can approximate the desired covariance matrix by increasing the number of pulses, each pulse bringing two additional free parameters that can be optimized \footnote{If one also allows to vary the modulation depth, one gains an additional degree of freedom for each pulse}. As an example, we give the result of a numerical optimization to achieve maximal squeezing in the center of mass mode of atoms 1 and 2. The initial pulse with area $\xi_Pt=0.5$ squeezes the relative motion of the first two atoms. The subsequent pulse is modulated dressing of the second atom and the auxiliary, with pulse area and relative phase treated as optimization parameters. The dynamics of the variances pertaining to the first two atoms in the resulting sequence are plotted in Fig.~\ref{optvariances}.

This final state also features squeezing in the relative motion of the two atoms, as well as squeezing of both oscillators individually. Additional constraints may be imposed into the optimization procedure, and can be satisfied to a desired precision by increasing the number of dressing pulses. 
At the end of the sequence, atoms 1 and 2 remain entangled with the auxiliary atom, and the results plotted in Fig.~\ref{optvariances} correspond to the state of the two atoms upon tracing over the state of the auxiliary atom. Note that since the center-of-mass of the entire system is not affected by the interactions, it remains in its vacuum state. The sum of the displacements of any two atoms can be squeezed by coupling them to an auxiliary system, but this results in entanglement of the atom pair with the remainder of the composite system. 

\begin{figure}
\includegraphics[width=0.8\columnwidth]{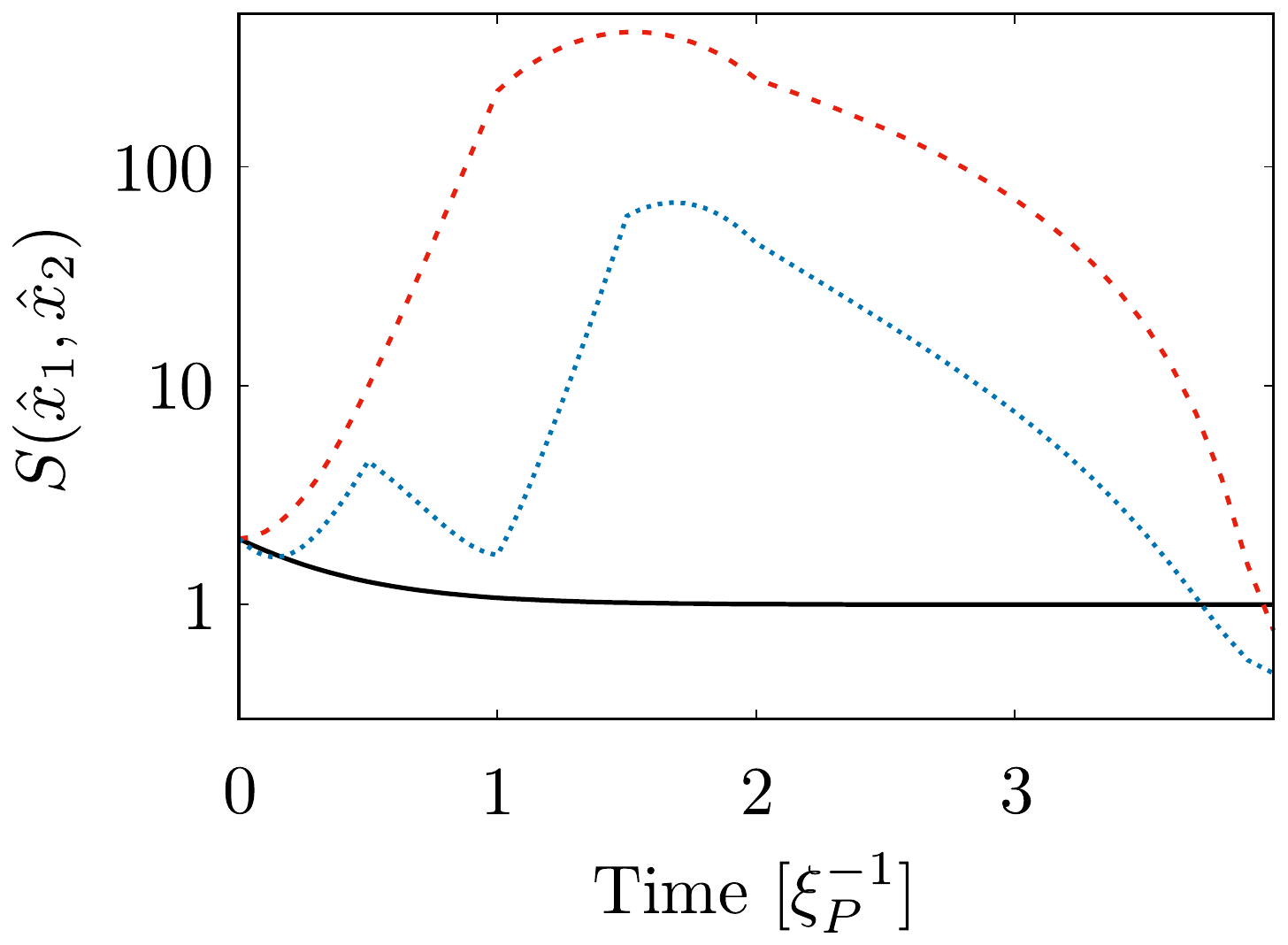}
\caption{Violation of the SPH criterion of the motional state of two atoms after applying a tailored sequence of pulses involving a single (solid, black), two (dashed, red) and three (dotted, blue) parametric amplifications in the pulse sequence.}
\label{simonplot}
\end{figure}

As a further example of the control in this system, we demonstrate how to maximize the non-classical correlations in the motional state of two selected atoms by a sequence of pulses. We may quantify the degree of entanglement between atoms 1 and 2 by the Simon-Peres-Horodecki (SPH) criterion~\cite{SPHCZ, epr}. Motional quadratures $\hat{x}_i$ and $\hat{p}_i$ with $[\hat{x}_i,\hat{p}_i]=i$ describing two {\it separable} oscillators satisfy
\begin{align}
S(\hat{x}_1,\hat{x}_2)\equiv\langle(\hat{x}_1+\hat{x}_2)^2\rangle+\langle(\hat{p}_1-\hat{p}_2)^2\rangle\ge 2.
\end{align}
Since  we can directly squeeze the relative motion, it is clear that $1\le S(\hat{x}_1,\hat{x}_2)\le 2$ can be achieved without the use of the auxiliary atom. Including an auxiliary system and more pulses allows to go beyond that. In Fig.~\ref{simonplot}, we give the value of $S(\hat{x}_1,\hat{x}_2)$ during the application of pulse sequences involving a single, two and three stages of parametric amplification. The parametric amplifications are applied between the target atoms (atoms 1 and 2), alternating with beam-splitter interactions between atom 2 and the auxiliary. A second parametric amplification stage allows $S(\hat{x}_1,\hat{x}_2)$ to reach a value as low as 0.75. Adding a second beam-splitter interaction and subsequent parametric amplification increases the final entanglement to $S(\hat{x}_1,\hat{x}_2)\le 0.5$ and also allows to reduce the intermediate phonon population of the entire system by an order of magnitude compared to a single parametric pulse. 

The control over the subsystem can be further improved by increasing the number of allowed pulses, increasing the number of atoms in the chain or re-cooling the auxiliary system. Alternatively, the state of the third atom can be taken into account rather than discarded. Appropriately tailored pulse sequences can generate motional many-body states useful as sensors for weak forces~\cite{dan} with high spatial resolution given by the different positions of the oscillators.

\subsection{State Transfer}
\label{transfer}
To illustrate the ability to distribute non-classical resources, we consider the problem of quantum state transfer along a chain of atoms. In more general terms, state transfer has been the subject of many theoretical \cite{StrRevs, nikolopoulos2014, nikolopoulos2004, Christandl2004, plenio2004, plenio2003} and experimental~\cite{transexp1, transexp2, transexp3} studies. Our aim is to create a motional state of a chain of atoms, which exhibits non-classical correlations between the first and last atom, while all the intermediate atoms are in their vacuum state. This is achieved by entangling the motion of the first pair of atoms and then transferring the quantum state of the second atom along the chain.  In a degenerate chain with identical nearest-neighbor interactions, perfect state transfer is possible only when there are at most two links in the chain. If the couplings between nearest neighbors can be tuned without affecting individual oscillator frequencies, there is a configuration of staggered couplings that allow perfect state transfer~\cite{transexp3, Christandl2004, nikolopoulos2004}. In our system, however, tuning the coupling between a given pair of atoms also influences the oscillation frequency of each atom individually, which precludes the perfect state transfer protocol~\cite{plenio2004}.

\begin{figure}[t]
\includegraphics[width=0.45\textwidth]{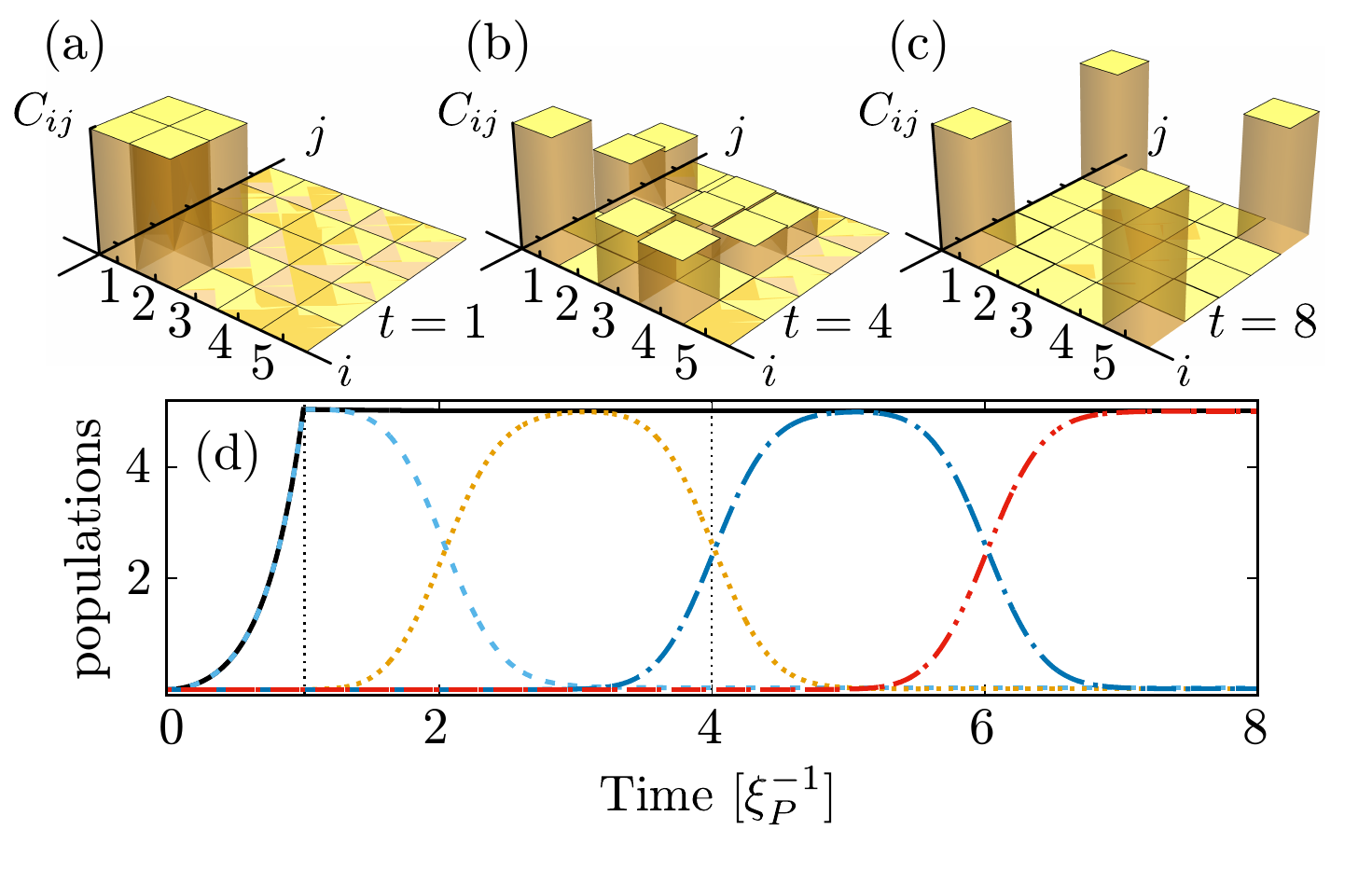}
\caption{Violation of the Cauchy-Schwarz inequality during the transfer sequence. (a), (b) and (c) give the value of $C_{ij}$, see Eq.~(\ref{cauchyschwarz}), after the modulated squeezing pulse, half way during the transfer and after the completion of the protocol respectively. (d) gives the excitation number $\langle \hat{a}^\dag_j\hat{a}_j\rangle$ in oscillator $j$ during the transfer. The dashed vertical lines give the times for plots (a) and (b).}
\label{transfig}
\end{figure}

Conceptually, the simplest way to achieve high-fidelity transfer across a long chain of oscillators coupled by springs is a sequence of swap operations between successive neighbors using the beam-splitter like interaction of Eq.~(\ref{reduceddyn}). A practical extension of this scheme is facilitated by our setup. In our system, the coupling between nearest-neighbors in a chain is mediated by a dressing laser irradiating both atoms. Sweeping a focused dressing laser beam across the chain naturally creates sequential links between subsequent atoms. If the speed of the sweep is chosen appropriately with respect to the beam width and strength, the dressing beam ``drags'' the quantum state of an atom across the chain. 
We simulate this transfer protocol for a chain of five atoms,  assuming a Gaussian transverse intensity profile for the dressing beam. The width of the beam is chosen to be half the separation between the neighboring atoms. We assume a sweep with constant rate and neglect the forces affecting coherent motion, see appendix ~\ref{appd} for details. This approximations is justified by the couplings $\xi_0$ being much smaller than the bare trapping frequencies $\omega_m$. The results of the simulation are plotted in Fig.~\ref{transfig} and show state transfer with nearly unit fidelity. 

Although all pairs of atoms are in principle coupled by the tails of the Gaussian beam profile, the transfer fidelity remains close to unity. The spread of correlations to more distant atoms in the chain -- which would decrease the transfer fidelity -- is naturally suppressed because only the two neighboring atoms near the focus of the dressing beam are resonant with each other at any given time.  

The proposed protocol is not unique, but uniquely versatile. The dynamics of the swept dressing beam can be optimized to distribute, e.g.,  more involved multipartite entanglement states along the system and the transfer protocol does not depend on the nature of the state to be transferred or on the length of the chain. The same protocol would equally well transfer and distribute non-classical features of a non-Gaussian state, such as a number state. The described setup can be realized in an optical lattice~\cite{pohl2014, pfau2014, bloch2016, saffman2013} or an array of optical microtraps~\cite{browaeys2004}. 
The ability to turn all interactions off after the completion of the transfer allows the preservation of the final state -- or any intermediate state for that matter -- for the long coherence time of atomic motion.

If the couplings are much stronger than motional decoherence times,  adiabatic state transfers become feasible. For such schemes the first and last atom in the chain are left degenerate, while all intermediate ones are coupled and far detuned. Such a setting will in almost all cases result in the state being transferred with high fidelity, albeit much slower than in the sequential protocol.

\section{Experimental considerations}

Throughout our treatment, we have neglected motional decoherence. Compared to the timescales of radiative relaxations, atomic motion in an ensemble of ultracold atoms is a long-lived degree of freedom. In our calculations for the interaction strengths, Fig.~\ref{interactionstrengths}, we assumed Rb atoms dressed with the $n=53$ Rydberg state. For the shortest lived $nS$ or $nP$ states, the radiative decay rate is $\gamma_r \sim 10\:$kHz \cite{numbers}, which, with the ratio $\Delta/\Omega \simeq 10$, leads to an estimate for the coherence time $\tau \simeq \gamma_r (\Delta/\Omega)^2 \sim 10\:$ms. A further source of decoherence is collisions with the background gas. Current experiments with trapped ultracold atoms have demonstrated motional coherence times in excess of 10 ms~\cite{spethmann2016}. For our parameters, leading to $\xi_1=\xi_0\simeq 1\:$kHz, we can estimate that the coherent atomic motion and its manipulation are not deteriorated significantly for times $t \lesssim 10/\xi_P$, which is not exceeded in the considered examples. The most effective experimental handle to increase motional couplings is to achieve higher Rabi-frequency $\Omega$ of the dressing field. Our proposal benefits from weak requirements on laser phase stability and linewidth, because the induced interatomic interactions depend on the field intensity. 

For the experimental verification of the described effects, it is necessary to measure the positions or momenta of individual atoms. Although not an easy task, recent developments in the field of quantum gas microscopy~\cite{greiner2015, zwierlein2015, takahashi2016,kuhr2015} have demonstrated sufficient sensitivities with reasonable statistical requirements.

Tighter atomic confinement decreases the effective strength of interatomic couplings due to the small spatial extent of the corresponding motional wave-function. Progress in Rydberg-dressing with high Rabi frequency could extend our protocols to tightly trapped atoms, which would allow the read-out via side-band detection techniques routinely used in trapped ion experiments~\cite{wineland1996}. These schemes allow the projection of atomic motion onto a Fock-state basis which would permit engineering the motional states beyond the Gaussian regime.

\section{Conclusions}

To summarize, 
we have investigated the possibility to manipulate the motional quantum state of a one-dimensional array of trapped neutral atoms via Rydberg-state dressing. Shining a dressing laser onto two neighboring ground-state atoms couples their motion through the interatomic forces they inherit from the van der Waals interaction between Rydberg-state atoms. Modulating the intensity of the dressing laser allows to squeeze the relative motion of two neighboring atoms and introduce non-classical correlations in the system. Expanding this toolbox to a chain of atoms, we demonstrated the possibility to access the center of mass mode of two atoms, create highly entangled states of motion and transfer non-classical correlations by sweeping the dressing laser with appropriate strength and speed along the chain. We used numerical parameters compatible with the current state-of-the art experiments and showed that our scheme can create non-classical states of motion within typical motional decoherence times. 

Further studies will include the expansion of the motional control to 2-dimensional arrays and the simulation of interesting many-body Hamiltonians using the motion of periodically driven lattices of neutral atoms. 

\begin{acknowledgments}
This work was supported by the H2020 FET Proactive project RySQ and by the Villum Foundation. LFB acknowledges fruitful discussions with Felix Motzoi.
\end{acknowledgments}

\newpage

\begin{appendix}

\section{Adiabatic Elimination}
\label{appa}
We work in the basis
\begin{align}
\Big\{|gg\rangle,|ge\rangle,|eg\rangle,|ee\rangle\Big\},
\end{align}
where Hamiltonian (\ref{hamiltonian}) acts as 
\begin{align}
H|gg\rangle=&\Omega(
|ge\rangle+|eg\rangle
)
\\
H|eg\rangle=&\Omega
|ee\rangle+\Omega|gg\rangle
-\Delta|eg\rangle
\\
H|ge\rangle=&\Omega
|gg\rangle+\Omega|ee\rangle
-\Delta|ge\rangle
\\
H|ee\rangle=&\Omega(|eg\rangle+|ge\rangle)-2\Delta|ee\rangle+V(R)|ee\rangle.
\end{align}
In terms of internal degrees of freedom, we write our wave function
\begin{align}
\psi=c_{gg}|gg\rangle+c_{ge}|ge\rangle+c_{eg}|eg\rangle+c_{ee}|ee\rangle.
\end{align}
Although this is not explicit here because it will not enter the elimination procedure, the coefficients depend in general on spatial coordinates.
We have the Schr\"odinger equation
\begin{align}
i\hbar\partial_t\psi=H\psi,
\end{align}
which gives us
\begin{align}
i\hbar\dot{c}_{gg}=&\Omega c_{eg}+\Omega c_{ge}
\\
i\hbar\dot{c}_{eg}=&\Omega c_{gg}-\Delta c_{eg}+\Omega c_{ee}
\\
i\hbar\dot{c}_{ge}=&\Omega c_{gg}-\Delta c_{ge}+\Omega c_{ee}
\\
i\hbar\dot{c}_{ee}=&\Omega c_{eg}+\Omega^* c_{ge}+(-2\Delta+V(R))c_{ee}
\end{align}
To eliminate the excited states, we set the time derivatives of $c_{eg},c_{ge}$ and $c_{ee}$ to zero, assuming that the dressing laser is not resonant with any of the transitions. We thus have
\begin{align}
i\hbar\dot{c}_{gg}=&\Omega (c_{eg}+ c_{ge})
\\
\Delta c_{eg}=&\Omega (c_{gg}+ c_{ee})
\\
\Delta c_{ge}=&\Omega (c_{gg}+ c_{ee})
\\
(2\Delta-V(R))c_{ee}=&\Omega(c_{eg}+ c_{ge}).
\end{align}
Adding the second and third equations leaves
\begin{align}
i\hbar\dot{c}_{gg}=&\Omega (c_{eg}+ c_{ge})
\\
\Delta(c_{eg}+c_{ge})=&2\Omega (c_{gg}+ c_{ee})
\\
(2\Delta-V(R))c_{ee}=&\Omega(c_{eg}+ c_{ge}),
\end{align}
which simplifies after some algebra to
\begin{align}
i\hbar\dot{c}_{gg}=& \left[ \frac{2\Omega^2}{\Delta}+\frac{4\Omega^4/\Delta}{2(\Delta^2-\Omega^2)-\Delta V(R)} \right] c_{gg}
\end{align}
From this expression we can read out the effective interaction potential (\ref{helim}). 

\section{Oscillating drives}
\label{appb}
We expand the dressing intensity (\ref{omegatimedep}) as 
\begin{align}\label{omegaexp}
|\Omega(t)|^4=|\bar{\Omega}|^4\sum_{j=0}^4 f_j\cos(j\omega_d t),
\end{align}
with 
\begin{subequations}\label{Aj}
\begin{align}
f_0=&1+3A^2+3/8A^4,
\label{omega0}\\
f_1=&4A+3A^3
\label{omega1}\\
f_2=&3A^2+1/2A^4,
\label{omega2}\\
f_3=&A^3,
\label{omega3}\\
f_4=&1/8A^4.
\label{omega4}
\end{align}
\end{subequations}
The oscillating term with the largest magnitude is (\ref{omega1}). Using this term for parametric driving requires $\omega_d=2(\omega_m-2\xi|\bar{\Omega}|^4)$.

\section{Analytic solution}
\label{appc}

The Heisenberg equations of motion are easily found from Hamiltonian (\ref{inthamiltonian}). In a new co-rotating frame, we have
\begin{align}
\frac{d\hat{\bf{a}}}{dt}=C\hat{\bf{a}},
\end{align}
with the matrix 
\begin{align}
C=
\left(\begin{array}{cccc}
0 & -i\xi_1e^{i\varphi} & -i\xi_0 & i\xi_1e^{i\varphi}\\
i\xi_1e^{-i\varphi} & 0 & -i\xi_1e^{-i\varphi} & i\xi_0 \\
-i\xi_0 & i\xi_1e^{i\varphi} & 0 & -i\xi_1e^{i\varphi} \\
-i\xi_1e^{-i\varphi} & i\xi_0 & i\xi_1e^{-i\varphi} & 0
\end{array}
\right)
\end{align}
The exponential of this matrix can be found analytically and has the block form
\begin{align}
e^{Ct}=\frac{1}{2}\left(
\begin{array}{cc}
C_+ & C_- \\
C_- & C_+
\end{array}
\right),
\end{align}
with 
\begin{align}
C_\pm=&
\left(
\begin{array}{cc}
e^{-it\xi_0}&0\\
0&e^{it\xi_0}\\
\end{array}
\right)
\nonumber\\
&\pm\left(\begin{array}{cc}
\cosh(\bar{\xi}t)+i\frac{\xi_0}{\bar{\xi}}\sinh(\bar{\xi}t)
& -2ie^{i\varphi}\frac{\xi_1}{\bar{\xi}}\sinh(\bar{\xi}t)
\\
2i\frac{\xi_1}{\bar{\xi}}e^{-i\varphi}\sinh(\bar{\xi}t)
&
\cosh(\bar{\xi}t)-i\frac{\xi_0}{\bar{\xi}}\sinh(\bar{\xi}t)
\end{array}
\right),
\end{align}
with $\bar{\xi}=\sqrt{4\xi_1^2-\xi_0^2}$. The transition towards parametric amplification occurs when $\bar{\xi}$ becomes real: $\xi_1>\xi_0/2$. Substituting in this inequality the expressions for $\xi_i$ involving $f_0$ and $f_1$, see Eqs. (\ref{Aj}),  yields the numerical values for the corresponding modulation depth in the main text. Transforming back to a frame rotating at $\omega_m$ yields the equations in the paper.

\section{State Transfer}
\label{appd}
Assuming a Gaussian intensity profile gives 
\begin{align}
|\Omega(t,x)|^2=I_0(t)e^{-\frac{(x-x_c(t))^2}{\sigma^2}}
\end{align}
where $x_c(t)=x_c(0)+vt$ denotes the position of the focus of the dressing laser beam, $\sigma$ is the width of the beam and $v$ is the speed at which the beam is dragged along the chain. For the interaction due to the dressing by this laser, we find 
\begin{align}
|\Omega(t)|^4\to|\Omega_i(t)|^4=I_0(t)^2e^{-\frac{(\langle x_i\rangle-x_c(t))^2}{\sigma^2}}e^{-\frac{(\langle x_{i+1}\rangle-x_c(t))^2}{\sigma^2}}.
\end{align}
For the state transfer in Fig.~\ref{transfig}, we used $\xi I_0(t)^2=9\xi_P$ and \mbox{$v=0.5R_0\xi_P$}.
\end{appendix}

\end{document}